\documentstyle[12pt,epsfig]{article}
\def\beq{\begin{equation}}   \def\eeq{\end{equation}}
\def\bea{\begin{eqnarray}}   \def\eea{\end{eqnarray}}

\newcommand{\bibit}[1]{\bibitem{#1}}

\newcommand{\gsim}{\lower.7ex\hbox{$ \;\stackrel{\textstyle>}{\sim}\;$}}
\newcommand{\lsim}{\lower.7ex\hbox{$ \;\stackrel{\textstyle<}{\sim}\;$}}

\newcommand{\matel}[3]{\langle #1|#2|#3\rangle}

\newcommand{\GeV}{\,\mbox{GeV}}

\begin{document}

\begin{flushright}
UND-HEP-00-BIG\hspace*{.2em}01\\
hep-ph/0005089\\
\end{flushright}
\vspace{.3cm}
\begin{center} \Large 
{\bf $D^0 \!-\! \bar D^0$ Oscillations as a Probe of  
Quark-Hadron Duality}\\
\end{center}
\vspace*{.3cm}
\begin{center} {\Large 
I. I. Bigi $^{a}$\\
N.G. Uraltsev $^{a,b}$}\\ 
\vspace{.4cm}
{\normalsize 
$^a${\it Physics Dept.,
Univ. of Notre Dame du
Lac, Notre Dame, IN 46556, U.S.A.} }
\\
$^b${\it Petersburg Nuclear Physics Inst., Gatchina, 
St.\,Petersburg, 188350, Russia} \\
\vspace{.3cm}
\vspace*{1.4cm}

{\Large{\bf Abstract}}\\
\end{center}

It is usually argued that the Standard Model predicts slow $D^0-\bar
D^0$ oscillations with $\Delta M_D,\, \Delta \Gamma _D \leq
10^{-3}\cdot \Gamma _D$ and that New Physics can reveal itself
through $\Delta M_D$ exceeding $10^{-3}\cdot \Gamma _D$.  It is believed 
that the bulk of the effect is due to long
distance dynamics that cannot be described at the quark level.
We point out that in
general the OPE yields soft GIM suppression scaling only
like $(m_s/\mu_{\rm hadr})^2$ and even like $m_s/\mu_{\rm hadr}$
rather than $m_s^4/m_c^4$ of the simple quark box diagram.  Such
contributions can actually yield $\Delta M_D,\, \Delta \Gamma _D
\sim {\cal O}(10^{-3})\cdot \Gamma _D$ without invoking additional
long distance effects. They are reasonably suppressed as long as the OPE 
and local duality are qualitatively applicable in the $1/m_c$ expansion.
We stress the importance of improving
the sensitivity on $\Delta \Gamma _D$ as well as $\Delta M_D$ in a
dedicated fashion as a laboratory for analyzing the onset of
quark-hadron duality and comment on the recent preliminary study
on $\Delta \Gamma _D$ by the FOCUS group.
 
\thispagestyle{empty}
\addtocounter{page}{-1}

\vspace*{.2cm}
\vfill
\noindent
\vskip 5mm

\newpage

$K^0 \!-\! \bar K^0$ oscillations have played a crucially important role
in the development of the Standard Model (SM). Likewise the
observation of $B_d \!- \!\bar B_d$ oscillations had an essential impact.
In both cases the oscillation rate is rather similar to the decay
rate. On the other hand one expects $D^0 \!-\! \bar D^0$ oscillations to
be very slow even on the scale of a second-order weak
amplitude. Searching for them thus represents primarily a high
sensitivity probe for New Physics in the electroweak sector with
almost zero background from SM dynamics.
Active searches are being undertaken using 
high quality data already in hand or soon to
become  available. 

In this note we look closely at the SM predictions. It is found 
that the operator product expansion includes contributions exhibiting
only very soft GIM suppression. They are much larger than the usual
quark box contribution and can dominate overall mixing.
We address quark-hadron duality 
(or duality for short), the role it plays here and what a signal or 
lack thereof can teach us about it and QCD. Such a
lesson is not merely of a conceptual nature, but can give us
information on the scale above which duality holds at
least approximately.

\section{SM Estimates of $D^0 - \bar D^0$ Oscillations}  

\subsection{General Features}

Oscillations in general 
are described by two dimensionless 
quantities: 
\beq 
x = \frac{\Delta M}{\bar \Gamma} \; , \;  
y = \frac{\Delta \Gamma}{2 \bar \Gamma} 
\eeq 
where 
\beq   
\Delta M \equiv M_2 - M_1 \; , \qquad  
\Delta \Gamma \equiv \Gamma _1 - \Gamma _2 \; , \;  
\bar \Gamma = \frac{1}{2} \left( \Gamma _1 + \Gamma _2 
\right) \; . 
\eeq  
The leading contributions to $\Delta M_K$ as well as 
$\Delta M_B$ are obtained from the well-known 
quark box diagram. The fields in the internal loop --  
$c$ in the former and 
$t$ quarks in the latter case
besides $W$ bosons -- are much heavier than the external 
quark fields. Those heavy degrees of freedom 
can be integrated out leading to  
$\Delta M_{K,B}$ being described by 
the expectation value of a local  
operator.\footnote{It turns out 
that $\Delta M_K$ is not completely 
dominated by local contributions reflecting dynamics 
operating around the scale $m_c$: a significant 
fraction is due to long distance dynamics characterized by 
low scales $\sim \Lambda_{\rm QCD}$. This would change 
dramatically if $m_c$ were larger.} This has been a highly 
successful ansatz: within the present 
theoretical uncertainties the SM can reproduce  
the observed values of $\Delta M_K$ and $\Delta M_{B_d}$ 
without forcing any parameter.

While the 
size of $\Delta \Gamma _K$ is naturally understood as due 
to $K\to 3 \pi$ being severely phase space restricted, 
it cannot be inferred from 
such a {\em local}  operator, as described above. 
For neutral $B$ mesons, on the other hand $\Delta \Gamma_B$ 
can be estimated through short distance dynamics 
\cite{DGBS}.

On very general grounds one expects 
$D^0- \bar D^0$ oscillations to be quite slow within the SM, 
since two structural reasons combine to make $x_D$ and $y_D$ 
small: 
\begin{itemize}
\item 
While the bulk of charm decays is 
Cabibbo allowed, the amplitude for 
$D^0 \!\leftrightarrow \!\bar D^0$ transitions is necessarily twice 
Cabibbo suppressed 
-- as is therefore the ratio between oscillation and decay 
rate:   
$\Delta M_D/\Gamma _D, \, 
\Delta \Gamma _D/\Gamma _D \propto  {\cal O}(\sin^2{\theta_C})$. 
The amplitudes for $K^0 \leftrightarrow \bar K^0$ 
and $B_d \leftrightarrow \bar B_d$ are Cabibbo and KM 
suppressed, respectively -- yet so are their decay widths 
allowing the oscillation and decay rate to be quite
comparable.
\item 
Due to the GIM mechanism one has $\Delta M=\Delta \Gamma =0$ 
in the limit of flavor symmetry. Yet flavor symmetry  
breaking driving  
$K^0 \to \bar K^0$ is characterized by $m_c^2 \neq m_u^2$  
and therefore no real suppression arises. 
On the other hand $SU(3)$ breaking 
controlling $D^0 \to \bar D^0$ is typified by $m_s^2 \neq m_d^2$
(or, in terms of hadrons, $M_K^2 \neq M_{\pi}^2$) as compared to the
scale $M_D^2$; 
it provides a significant reduction. 
\end{itemize}
Having two Cabibbo suppressed classes of decays 
one concludes on these very
general grounds:\footnote{One can argue that because of the 
$\Delta C = - \Delta Q$ rule in semileptonic charm decays one should 
write the nonleptonic rather than the total width in 
Eq.\,(\ref{BOUNDGEN}); yet this difference is small for $D^0$ 
and certainly is in the theoretical `noise'.} 
\beq 
\Delta M_D\; , \; \Delta \Gamma _D 
\sim \; SU(3) \; {\rm breaking} \; 
\times 2 \sin^2{\!\theta_C} \times \Gamma _D 
\label{BOUNDGEN}
\eeq
The proper description of $SU(3)$ {\it breaking} thus becomes the 
central issue.

With typical nonleptonic $D$ decay channels exhibiting sizeable 
$SU(3)$ breaking  -- see our discussion in Sect.\,2 -- 
{\it a priori} one cannot count on this suppression to amount to more
than a factor of about two or three in the $D$ {\it width
difference}. There are reasons to believe that 
a larger reduction 
may occur for the mass difference $\Delta M_D$ driven by virtual
intermediate states. Yet an order of magnitude reduction, in
particular in $\Delta\Gamma_D$ would seem unjustifiably
pessimistic. Thus 
\beq 
\frac{\Delta M_D}{\Gamma _D} \; \lsim \;  
\frac{\Delta \Gamma _D}{\Gamma _D}  
\lsim \;\frac{1}{3} \; 
\times 2 \sin^2{\!\theta_C} \;  \sim \;
{\rm few} \times 0.01
\label{GENERAL} 
\eeq
represents a conservative bound for overall mixing based on 
very general features of the SM; 
for the mass difference this estimate can 
actually be seen on the cautious side.

The following line of arguments is usually employed: 
({\bf i}) Quark-level contributions are estimated  
by the usual quark box diagrams and yield only insignificant contributions 
to $\Delta M_D$ and $\Delta \Gamma _D$ (see below). 
({\bf ii}) Various schemes employing contributions of selected
hadronic states are invoked to estimate 
the impact of long distance 
dynamics; the numbers typically resulting are 
$x_D \, , \; y_D \,\sim \, 10^{-4} - 10^{-3}$ 
\cite{BURD,apetrov,BUCCELLA}.  
({\bf i\hspace*{-.2mm}i\hspace*{-.2mm}i}) These findings 
lead to the following widely 
embraced conclusions:  
An observation of $x_D > 10^{-3}$ would reveal the intervention of New 
Physics beyond the SM,   
while $y_D \simeq \left. y_D \right|_{SM} \leq 10^{-3}$ has 
to hold since New Physics has hardly a chance to contribute to it. 

Beyond the general property that both $\Delta M_D$ and 
$\Delta \Gamma _D$ have to vanish in the $SU(3)$ limit, the dynamics 
underlying them have different features: 
$\Delta M_D$ receives contributions from virtual  
intermediate states whereas $\Delta \Gamma$ is generated by 
on-shell transitions. Therefore the former is usually considered 
to represent a more robust quantity than the latter; actually it 
has often been argued that quark diagrams {\em cannot} be relied upon 
to even estimate $\Delta \Gamma$. A folklore has arisen that theoretical 
evaluations of the two quantities rest on radically different grounds.  
 
Yet we note that despite these 
differences there is no fundamental distinction  
in the theoretical treatment of $\Delta M_D$ and $\Delta \Gamma _D$:  
both can be described through an operator product expansion, and
its application relies on local quark-hadron duality for both 
$\Delta M_D$ and $\Delta \Gamma _D$. Only the 
numerical aspects differ, as does the sensitivity to New Physics.

\subsection{Operator Product Expansion}

Following the general treatment of inclusive weak transitions, 
see Refs.\,\cite{rev,inst,d2}, we can describe 
$D^0 \!-\! \bar D^0$ oscillations 
by considering a correlator 
\beq 
\hat T_{D\bar D}(\omega) \!=\! \,\mbox{\small $\frac{1}{2}$}
 \int {\rm d}^4x\, {\rm e}\,^{-i\omega t}\:
iT \{{\cal H}_W(x) {\cal H}_W(0)\}\;, \;\;\; 
T_{D\bar D}(\omega) \!= \!\frac{1}{2M_D} 
\matel{\bar D}{\hat T_{D\bar D}(\omega)} {D}
\label{A2}
\eeq
as a function of a complex variable $\omega$. Here 
${\cal H}_W$ is the $\Delta C \!=\!-1$ Hamiltonian density.
With the mixing amplitude of interest 
\beq 
A(\omega)= 2 \sum_n \frac{1}{2M_D} 
\frac{\matel{\bar D}{{\cal H}_W}{n(\vec{k}\!=\!0)}
\,
\matel{n(\vec{k}\!=\!0)}{{\cal H}_W}{D}}
{E_n\!-\!M_D\!+\!\omega \!+\!i\epsilon }
\equiv -\Delta \tilde M_D(\omega) +\frac{i}{2}\Delta \tilde \Gamma_D(\omega)
\label{A3}
\eeq
one has $4\,T_{D\bar D}(\omega)\!=\! A(\omega)\!+ \!A(\!-\!\omega)$,  
whereas $\Delta M_D\!=\!\Delta \tilde M_D(0)$, 
$\Delta \Gamma_D\!=\!\Delta \tilde \Gamma_D(0)$.    
The summation above runs over all intermediate states $|n \rangle$
with energies $E_n$ and vanishing spacelike momentum. 
$\Delta M_D$ can be expressed by a dispersive integral 
over $\Delta \tilde
\Gamma_D(\omega)$ evaluated through the 
principal value prescription: 
\beq
\Delta M_D = \frac{1}{2\pi} \:{\rm V\!\!.P\!\!.} 
\int {\rm d}\omega\: \frac{\Delta \tilde\Gamma_D(\omega)}{\omega}
\;.
\label{A6}
\eeq

Applying the operator product expansion (OPE) to 
Eq.\,(\ref{A2}) provides us with a consistent evaluation of the 
transition rates through an expansion in powers of 
$1/m_c$. With the charm quark mass exceeding the typical 
scale of strong interactions $\mu_{\rm hadr}$ by 
a modest amount only, one cannot 
count on obtaining a reliable quantitative description in this way; 
yet it still yields a useful classification of various 
effects. This is briefly reviewed below.
\vspace*{.2cm}

The leading term for $\Delta C\!=\!-2$ transitions comes from dimension-6
four-fermion operators of the generic form $(\bar{u}c)(\bar{u}c)$ 
with the  
corresponding Wilson coefficient receiving contributions from 
different sources. 

({\bf a}) Effects due to intermediate $b$ quarks are most simply 
calculated since
they are highly virtual and Euclidean:
\beq
\Delta M_D^{(b \bar b)} \simeq  \frac{G_F^2 m_b^2}{8\pi ^2} 
\left| V^*_{cb}V_{ub}\right| ^2 \frac{1}{2M_D}  
\matel{\bar D^0}{(\bar u \gamma_{\mu} (1 \!-\! \gamma _5)c)
(\bar u \gamma_{\mu} (1 \!- \!\gamma _5)c)}{D^0}\; ;  
\label{B2}
\eeq
however they are highly suppressed by
the tiny KM mixing with the third generation. Using factorization to
estimate the  matrix element one finds: 
\beq
x_D^{(b\bar b)} \sim \:{\rm few} \times 10^{-7}\;.
\label{B3}
\eeq
Loops with one $b$ and one light quark likewise are suppressed.

({\bf b})
For the light intermediate quarks the momentum scale is set by the
external mass $m_c$, and the corresponding factor is given by $G_F^2
m_c^2/8\pi^2 \sin^2{\!\theta_C}\cos^2{\!\theta_C} $ 
(from now on we will often omit the KM factors when they are
obvious). However, it is highly suppressed by the GIM factor 
$\left(\frac{m_s^2\!-\!m_d^2}{m_c^2}\right)^2$ 
leading to 
\footnote{ 
This contribution is obviously saturated at the momentum scale
$\sim \!m_c$, and thus refers to the Wilson coefficient of the $D\!=\!6$
operator. We disagree with statements that these are
long-distance contribution simply 
because they are proportional to $m_s^2$.} 
$$
\hat{T}_{D\bar{D}}(\omega)=
 \frac{G_F^2}{16\pi^2} 
\left| V^*_{cs}V_{us}\right| ^2 
\left(
\frac{\left( m_s^2 \!-\! m_d^2\right) ^2}{(m_c-\omega)^2} + 
\frac{\left( m_s^2 \!-\! m_d^2\right) ^2}{(m_c+\omega)^2}
\right) \;\times \qquad\qquad\qquad\qquad\qquad
$$\
\beq
\qquad\qquad\qquad\qquad
\left[
(\bar u \gamma_{\mu} (1\!-\!\gamma_5)c)
(\bar u \gamma_{\mu} (1 \!-\!\gamma _5)c) + 
2(\bar u (1\!+\!\gamma _5)c)
(\bar u  (1\!+\!\gamma _5)c)\right]\;.
\label{B3a}
\eeq 
 Hence we read off 
for its contribution to the mass
difference
$$  
\Delta M_D^{(s,d)} \simeq   
- \frac{G_F^2 m_c^2}{4\pi ^2} 
\left| V^*_{cs}V_{us}\right| ^2 
\frac{\left( m_s^2 \!-\! m_d^2\right) ^2}{m_c^4} 
$$ 
\beq 
\times \:
\frac{1}{2M_D}\matel{\bar D^0}{(\bar u \gamma_{\mu} (1 \!-\! \gamma _5)c)
(\bar u \gamma_{\mu} (1 \!-\! \gamma _5)c) + 
2(\bar u (1 \!+ \!\gamma _5)c)
(\bar u  (1 \!+ \!\gamma _5)c)}{D^0}
\;.
\label{B4}
\eeq
(This expression differs from what is usually quoted in the
literature, see, e.g.\ Ref.\,\cite{BURD}.)
As follows from Eq.\,(\ref{B3a}) bare quark loops do not contribute
to $\Delta \Gamma_D$ at this order.  The latter is 
suppressed by additional powers of $m_s/m_c$, or by $\alpha_s/\pi$ when
gluon corrections are accounted for (e.g., through the anomalous 
dimension of the light quark mass).

The GIM suppression by two powers of $m_s/m_c$ for each quark 
line is inevitable for left-handed weak vertices. This feature 
persists for Penguin operators, albeit in a slightly
different way. Numerically one finds: 
\beq 
\Delta \Gamma_D^{\rm box} < 
\Delta M_D^{\rm box} \sim {\rm few} \; \times 10^{-17} \; {\rm GeV} 
\; \; \hat = \; \; x_D^{\rm box} \sim {\rm few} \; \times 10^{-5} 
\label{B6} 
\eeq  
\vspace*{.2cm}

However, since the leading Wilson coefficient is highly suppressed,
one has to consider also the contributions from higher dimensional
operators. It turns out that the $SU(3)$ GIM suppression is in
general not as severe as $(m_s^2 \!-\! m_d^2)/m_c^2$ per fermion line: it
can be merely $m_s/\mu_{\rm hadr}$ if the fermion line is soft. In
the so-called practical version of the OPE
\cite{OPE} this is described by condensates
contributing to the next terms in the $1/m_c$ expansion. 

There is a simple rule of thumb: cutting a quark line, we
pay the price of a power suppression $\sim \mu_{\rm hadr}^3/m_c^3$; yet
the  GIM suppression now becomes only $m_s/\mu_{\rm hadr}$. Altogether
this yields a factor $\sim 4\pi^2\mu_{\rm hadr}^2/(m_s m_c)$ which can
result in an enhancement. In particular, we keep in mind that $SU(3)$ 
breaking effects in condensates are not significantly suppressed, and
the ratio between $\mu_{\rm hadr}$ and $m_c$ is not much smaller than unity. 

\thispagestyle{plain}
\begin{figure}[hhh]
\begin{center}
\mbox{\psfig{file=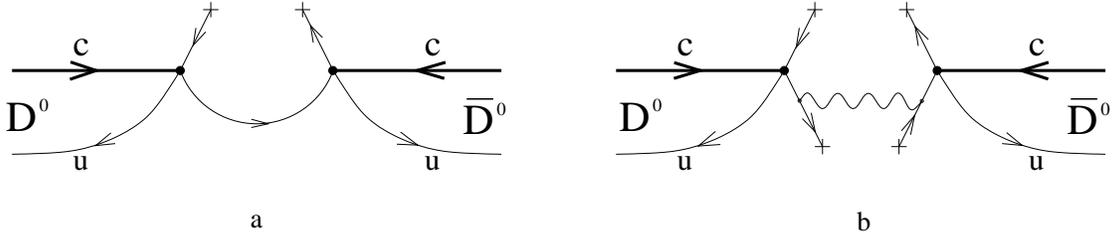,height=1.22in,width=5.8in}}
\end{center}
\caption{Diagrams generating higher-dimension operators suffering 
softer GIM suppression}
\end{figure}

An example is given by the diagram in Fig.\,1a. It yields a six-fermion
operator of the generic form
$
(\bar{u}c)(\bar{u}c)(\bar{d}d\!-\!\bar{s}s)
$  
with the Wilson coefficient 
$
\sim \frac{m_s^2}{m_c^2} \, G_F^2 m_c^{-1}\;.
$
This contribution thus scales like 
$4\pi^2 m_s^3 \mu_{\rm hadr}^3/m_c^5$ compared to the ``standard'' 
factor $G_F^2 f_D^2
m_c^2 M_D$. Explicitly, this contribution (neglecting gluon
corrections) is given by
\beq
\Delta M_D^{(D=9)}= 2\sin^2{\!\theta_C} \cos^2{\!\theta_C}\,
\frac{G_F^2 m_s^2}{m_c^3}\;\times \qquad \qquad\qquad\qquad\qquad
\label{B8}
\eeq
\vspace*{-.5cm}
$$ 
\frac{1}{2M_D}\matel{\bar D}{
\bar{u}^i\gamma_\mu(1\!-\!\gamma_5)c^j\, 
\bar{u}^k\gamma_\nu(1\!-\!\gamma_5)\gamma_0c^i\,
\left(
\bar{s}^j\gamma^\mu\gamma_0\gamma^\nu (1\!-\!\gamma_5)s^k -
\bar{d}^j\gamma^\mu\gamma_0\gamma^\nu (1\!-\!\gamma_5)d^k
\right)
}{D} 
\;.
$$ Here $i,j,k$ are color indices.  Note, that the $c$ quark
operators here are normalized at the low momentum scale. Therefore,
$\frac{1\!+\!\gamma_0}{2} c(x)$ describes only annihilation of $c$
quark, and $\frac{1\!-\!\gamma_0}{2} c(x)$ only creation of charmed
antiquark. This implies, for example, that the relation 
$\gamma_0 c(x)\, \times\, c(y) = \!-\: c(x) \,\times \, \gamma_0 c(y)$ 
always holds for the considered matrix elements, since only two
combinations $\frac{1 \pm \gamma_0}{2} c(x) \times \frac{1 \mp
\gamma_0}{2} c(y)$ survive. It is interesting that for the ``neutral
current'' type color flow in the both weak vertices, the
contributions of Figs.~1a (proportional to $a_2^2$) vanish to the
leading order in $1/m_c$.

\vspace*{.2cm}

$SU(3)$ suppression can be further softened by cutting both fermion
lines. To transfer a large momentum one has to add a gluon, like in
Fig.\,1b (for this reason another loop factor of $4\pi^2$ is
replaced by $4\pi\alpha_s$). These yield eight-fermion operators with 
the flavor structure 
\footnote{There are actually additional contributions not obtained 
merely by cutting quark lines; they are reminiscent of the soft part 
of Penguin contributions, see below.} 
$\,
(\bar{u}c)(\bar{u}c)\left[
(\bar{d}d)(\bar{d}d) + (\bar{s}s)(\bar{s}s)-
(\bar{d}d)(\bar{s}s)-(\bar{s}s)(\bar{d}d)   
\right].
$
With the $SU(3)$ suppression in the matrix element due to double
antisymmetrization between $s$ and $d$ only $m_s^2/\mu_{\rm hadr}^2$, 
this contribution scales like  
$\frac{4\pi^2 m_s^2 \mu_{\rm hadr}^4}{m_c^6}\cdot G_F^2 f_D^2
m_c^2 M_D$.

It is interesting to note that, in principle, the $SU(3)$ suppression
can be as mild as only the first power of $m_s$. Namely, if we
schematically define
$$
\matel{\bar D}{
(\bar{u}c)(\bar{u}c)\!\left[
(\bar{d}d)(\bar{d}d) \!+\! (\bar{s}s)(\bar{s}s) \!-\!
(\bar{d}d)(\bar{s}s)\! -\!(\bar{s}s)(\bar{d}d)   
\right]\!}
{D} \equiv \zeta \:
\matel{\bar D}{
(\bar{u}c)(\bar{u}c)(\bar{d}d)(\bar{d}d) 
}{D}
$$
then, at small $m_s$, the $SU(3)$ suppression factor $\zeta$ can
scale as $m_s/\mu_{\rm hadr}$. Indeed, if the matrix element, as a
function of the two quark masses is given by 
$$
\matel{\bar D}{
(\bar{u}c)(\bar{u}c)(\bar{q}_1q_1)(\bar{q}_2 q_2) 
}{D} = A + B\cdot(m_1\!+\!m_2) \ln{\frac{\mu_{\rm hadr}}{m_1\!+\!m_2}}
$$
then $\zeta \simeq -2B m_s \ln{2}$. More accurately, for the actual
operator the leading, linear
in $m_s$ term in the ``soft GIM'' factor $\zeta$ is
determined by the matrix element
\beq
\frac{M_K^2}{16\pi^2 f_\pi^2} <\bar D^0| (\bar{u}_L\Gamma c_L)^2 
\:\bar{d}_L \Gamma 
s_L  \,
\bar{s}_L \Gamma d_L| D^0>
\label{B12}
\eeq
in the limit of massless $s$ and $d$ and is due to the chiral
$\bar{K}^0\eta(\pi^0) K^0$ loop. The explicit structure of the
Lorentz and color matrices $\Gamma$ above follows from the operator
given below in Eq.\,(\ref{B14}). 

\thispagestyle{plain}
\begin{figure}[hhh]
\begin{center}
\mbox{\psfig{file=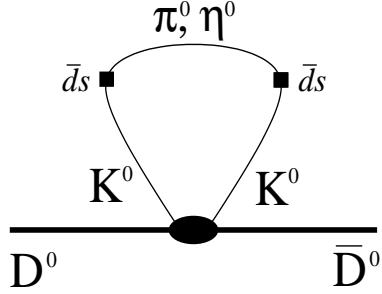,height=1.5in,width=2in}}
\end{center}
\caption{Chiral loop leading to the effect linear in $m_s$.}
\end{figure}

Let us sketch this step. At small $m_s$ (for simplicity we put $m_d\!=\!0$, 
but $m_u$ can be arbitrary) it is convenient to use the {\it current} 
quark fields $s'$, $d'$ instead of mass eigenstates, and perform an 
expansion around the symmetry point $m_s\!=\!m_d$. Then the transition
operator has the simple flavor structure $(\bar{u}c)^2
(\bar{s}'d')^2$. Its matrix element between $\bar{D}^0$ 
and $D^0$ vanishes to zeroth order 
since the operator has $\Delta
S'\!=\!-\Delta D'\!=\!-2$. 
The mass perturbation, however, has a $\Delta
S'\!=\!-\Delta D'\!=\!1$ piece $\sin{\theta_C}\cos{\theta_C}\,m_s\,
\bar{d}'s'$. The nonvanishing second-order correction to the matrix
element 
$\matel{\bar{D}^0}{(\bar{u}c)^2 (\bar{s}'d')^2}{D^0}$ 
is then generally proportional to $m_s^2$. The only exception comes
from the pseudo-goldstone loop of Fig.\,2 shaped by 
$$
\int \frac{{\rm d}^4 k}{(2\pi)^4 i}\, \frac{1}{(M^2-k^2)^3} \,=\, 
\frac{1}{32\pi^2 M^2}
$$
with $M^2 \propto m_s$. This contribution is proportional to the
zero-momentum amplitude 
$\matel{\bar{D}^0 \bar K^0 \bar K^0 }{(\bar{u}c)^2 (\bar{s}d)^2}{D^0}$
which, by PCAC is related to the chiral limit matrix element of the
double commutator of the operator with the $\bar{d}{s}$ axial
charge. This yields the stated equation.

The above estimate serves only as an {\it existence proof}. 
Most probably, such
infrared effects involving pion loops are not the dominant
source. Even without such nonanalytic terms the matrix elements
typically depend strongly on the quark masses, and in the actual
world the double subtraction present in the above operators, can
result in only a mild suppression factor in spite of being formally
of the order of $m_s^2$.  
\vspace*{.2cm}

Estimates of the actual size of these contributions  
at present suffer from considerable uncertainties, primarily 
in the matrix elements. 
Direct computation of the bare diagram yields the
following cumbersome result:
\beq
-2\, \hat{T}(0) 
= -\frac{8\pi\alpha_s G_F^2}{m_c^4} \sin^2{\!\theta_C} 
\cos^2{\!\theta_C}\;\, 
{\cal G}\, O^{(12)}\;,
\label{B14}
\eeq
$$
 O^{(12)} \;=\; \left\{
\bar{u}\Gamma^\alpha d\, \bar{s}\Gamma_0 t^a c -
 \bar{u}\Gamma_0 d\, \bar{s}\Gamma^\alpha t^a c 
+i\epsilon^{0\mu\alpha\beta}\,\bar{u}\Gamma_\beta t^a d\, \bar{s}\Gamma_\mu c
\right\}\;\times
\qquad\qquad\qquad\qquad\;\;\;
$$
$$
\qquad\qquad\qquad\qquad\qquad\qquad\qquad
\left\{\bar{u}\Gamma_\alpha s\, \bar{d}\Gamma_0 t^a c -
 \bar{u}\Gamma_0 s\, \bar{d}\Gamma_\alpha t^a c 
+i\epsilon_{0\nu\alpha\gamma}\,\bar{u}\Gamma^\gamma t^a s\, 
\bar{d}\Gamma^\nu c 
\right\}
$$
where $\Gamma_\mu \!=\! \gamma_\mu(1\!-\!\gamma_5)$ and
$t^a\!=\!\frac{\lambda^a}{2}$ are color matrices. The symbol ${\cal
G}$ denotes GIM-type subtraction; its action on a generic 
operator with four $s$ and $d$ quarks like in $O^{(12)}$ is defined as
\beq
{\cal G}\, (\bar{s} U d) (\bar{d} V s) \;= \qquad \qquad\qquad
\qquad\qquad\qquad
\label{G}
\eeq
$$ 
\; (\bar{s} U d) (\bar{d} V s)-(\bar{s} U s) (\bar{s} V s)-
(\bar{d} U d) (\bar{d} V d)+(\bar{d} U s) (\bar{s} V d) +
(\bar{s} U s) (\bar{d} V d)+(\bar{d} U d) (\bar{s} V s)
$$
(the last two terms do not have counterparts in the box diagram).
The exact
coefficients and the color structure in $O^{(12)}$ are 
modified by straightforward renormalization of the
weak decay operators, yet the major uncertainty lies in the matrix
elements.

In the spirit 
of the ``Educated dimensional analysis'' 
of Ref.\,\cite{GGG} we estimate the magnitude of the matrix element of 
${\cal G}\,O^{(12)}$ as
\beq
\zeta f_D^2 M_D \cdot (0.3\GeV)^6  \approx \zeta\cdot 
7\cdot 10^{-5}\GeV^9
\label{B16}
\eeq
where $\zeta$ accounts for the $SU(3)$ GIM suppression,
$$
\matel{\bar{D^0}}{{\cal G}\,O^{(12)}}{D^0} = 
\zeta \matel{\bar{D^0}}{O^{(12)}}{D^0}\;.
$$
Note that $\alpha_s$ 
enters at the charm mass scale, and for consistency must be evaluated in
the $V$- rather than the $\overline{\rm MS}$-scheme, which is obvious in
the BLM approximation. Numerically we end up with 
\beq
\delta^{(12)} x_D  \sim \;  {\cal O}(10^{-3})\;.
\label{B18}
\eeq
It is actually conceivable that medium-size instantons yield enhanced
contributions to the corresponding matrix element of the
eight-fermion operators in question, due to the induced t'~Hooft
vertex of the form $(\bar{u}_L u_R) (\bar{d}_L d_R) (\bar{s}_L s_R)$.
Note that it can yield the effect $\propto m_s^2$
which is not formally related to the spontaneous symmetry breaking
since would violate only the anomalous singlet $U_A(1)$. 

Diagrams in Figs.\ (1) and (2) literally do not produce an absorptive
part and, therefore contribute to $\Delta M_D$, but not $\Delta
\Gamma_D$. Yet the latter can be generated, for example, through a
cut across the gluon propagator in Fig.\,2 if it is dressed;  in
$\Delta M_D$ this would contribute to the anomalous dimension.
This is the
leading contribution in the BLM approximation. 
It amounts to replacing
$\alpha_s$ by $\mbox{{\small $\frac{9}{4}$\,}} 
i\alpha_s^2/(1+81/16\alpha_s^2)$ in
Eq.\,(\ref{B14}). Such approximation is justified if other
contributions to the anomalous dimension can be neglected.
At the charm scale
these modifications do not seem to lead to a particular numerical
suppression of $\Delta \Gamma_D$ compared to $\Delta M_D$. Therefore, 
we arrive at 
\beq 
x_D \; , \; \; y_D \; \;  \sim \; {\cal O}(10^{-3}) 
\label{B19}
\eeq 
\vspace*{.2cm}

{\em In summary:} We have shown that the high degree of $SU(3)$ invariance
and related GIM suppression $(m_s^2 - m_d^2)^2/m_c^2$ exhibited by
quark box diagrams for $D^0 \to \bar D^0$ is {\it not} typical for
the process.  
Terms in the $D^0 \to \bar D^0$ amplitude can be proportional to
$ m_s^2$ or even $m_s^1$. Such contributions arise naturally
in the OPE through condensate contributions containing higher
dimensional operators. While those are formally suppressed by powers
in the heavy quark mass, this does not constitute a very significant
factor for the case of charm.  It had been noted long ago (see, for
example, Ref.\,\cite{au}) that estimates of the absorptive part
of the $D^0 \to \bar D^0$ amplitude are very sensitive to low energy
parameters: evaluating it on the quark level one encountered much more
effective $SU(3)$ cancellations than in its potential hadronic
counterparts $\propto M_K^4/m_c^4$. Application of the OPE treatment allows
to clarify and justify those, rather tentative suggestions
identifying the possible source of such enhanced contributions in the 
framework of the $1/m_c$ expansion. Moreover, it becomes clear that
the same applies to the mass difference $\Delta M_D$ as well. 

Our numerical estimates are rather
uncertain.  However we note that the OPE naturally accommodates the
size of mixing previously discussed in the literature as a possible
effect of general long-distance dynamics.  At the same time, we think
that $x_D$, $y_D$ exceeding $5\cdot 10^{-3}$ cannot be
attributed to the OPE contributions in the framework of standard
assumptions.

Regardless of the size of the matrix elements involved, we still can
state that $D^0\!-\bar{D}^0$ mixing must be suppressed in the SM
whenever it makes sense to speak of it in the $1/m_c$ expansion. The
natural yardstick for the unsuppressed level is the overall
nonleptonic decay width (up to inherent CKM mixing factors). As
illustrated above, smallness of the higher-order terms compared to the 
formally leading in $1/m_c$ effect for mixing cannot serve as a valid 
universal criterion. Yet a certain suppression of the higher-order
terms compared to the (parton) decay width {\it per se} is a
necessary condition for applying sensibly the $1/m_c$
expansion. Since the identified effects with mild GIM cancellation
emerge in relatively high order in $1/m_c$, the minimal suppression
must amount to a noticeable factor, as asserted above.

\subsection{Quark-Hadron Duality}

Beyond the question about the size of the higher order corrections in
the $1/m_c$ expansion there is the more fundamental one about local
quark-hadron duality; i.e., to which degree of accuracy does a quark
level result derived from the OPE describe an inclusive quantity
involving hadrons?  In other words, how well knowledge of
$\Delta\tilde M(\omega)$, $\Delta\tilde \Gamma(\omega)$ in
Eq.\,(\ref{A3}) at large (compared to $ \Lambda_{\rm QCD}$) complex
$\omega$ based on the short-distance expansion, determines their
values at $\omega\!=\!0$ measured in experiment.

It was pointed out in Ref.\,\cite{optical} that with heavy quarks one
often has to deal with a novel aspect, referred to 
as {\em global} duality:
a Euclidean dispersive integral will reproduce the 
contributions coming from {\em all} cuts in the Minkowski domain,
while some of them are unphysical for the considered decay process. 
One then has 
to filter out the 
contributions to the integral that correspond to the individual process 
of interest. Since this can be done only to a certain accuracy, 
it introduces a further source of theoretical uncertainty. 

Even though the cuts in $T_{D\bar{D}}(\omega)$ describe the same
physical channel, this complication still persists in a certain form
for flavor oscillation processes: with
$T_{D\bar{D}}(\omega)\!=\! A(\omega)\!+\!
A(\!-\!\omega)$, it has two cuts which overlap. However, since one
of them starts at $\omega \simeq -m_c$ and another at $\omega
\simeq m_c$, they can be disentangled at $|\omega| \ll m_c$ in the
$1/m_c$ expansion of `practical' OPE in the same way as in the usual heavy
quark decay widths \cite{optical}. The central issue is again the
validity of local duality, both for $\Delta \Gamma$ and $\Delta
M$. Its more dedicated explanation and its implementations can be
found in various reviews (see, for example, Ref.\,\cite{rev,inst,d2}).

That also $\Delta M_D$ is sensitive to duality violations is readily
seen by imagining the presence of a narrow resonance of appropriate
quantum numbers close to the $D$ meson mass: it would significantly
affect the size of $\Delta M_D$. Yet, practically, it is natural to
expect local duality violations to be smaller in $\Delta M_D$ than in
$\Delta \Gamma _D$; i.e., the onset of duality should occur at a
lower scale for the former than for the latter.  For $\Delta \Gamma
_D$ is directly given by the discontinuity in the corresponding $D
\rightarrow \bar D$ transition amplitude, whereas $\Delta M_D$ can be
represented by the principal value of the dispersion integral,
Eq.\,(\ref{A6}). The latter provides a measure of averaging that 
reduces the sensitivity to the resonances or thresholds and thus local
duality violations. This has been illustrated by a simple model with
a single resonance in Ref.\,\cite{apetrov}. 

While no general proof has been given for the validity of 
local duality, considerable evidence has been 
accumulated over the years that it does apply for sufficiently 
heavy flavors. Detailed studies of OPE and duality have been 
performed 
recently within model field theories, in particular the 
t'~Hooft model 
\cite{THOOFT} which is QCD in 1+1 dimensions with 
$N_c \to \infty$. 
Analytical analyses showed that the OPE in the inverse powers of heavy
quark mass holds for the heavy quark decay
widths \cite{d2} in spite of certain doubts which had been
voiced. 
The above papers did not consider the 
width difference between the two neutral heavy meson eigenstates. Yet 
using the technique developed there, it is not difficult to
establish the similar correspondence between the hadronic saturation
and the quark box diagrams at least through the next-to-leading
order in $1/m_Q$.

Concluding duality to be valid asymptotically 
-- for $m_Q \to \infty$ -- one turns to the question 
at how low a scale duality emerges to apply with some accuracy. 
Most authors would expect it to be valid for $m_Q \sim m_b$; 
yet assuming duality to hold already at 
the charm scale even in a semiquantitative fashion would 
appear to be a rather iffy proposition for semileptonic 
transitions, let alone for nonleptonic ones. As argued above,
experimental observation of a stronger suppression of $D^0\!-\!\bar D^0$
oscillations compared to the phenomenological estimate
Eq.(\ref{GENERAL}), would be an indication of a relatively low onset of
duality for the inclusive decay widths. The {\em width} difference 
$\Delta \Gamma_D$ is an even more sensitive,  undiluted probe for duality 
violations than $\Delta M_D$. 
A scenario with a sizeable  
$\Delta \Gamma _D$ and a somewhat smaller value of $\Delta M_D$ 
could still imply the nearby onset of local duality. 

One reservation has to be made though, due to a notorious complication
peculiar to local duality violation. Since the duality violating 
component `oscillates' (as a function of $m_c$), it can actually 
vanish for certain mass values. Determining the size of 
such effects at a single scale 
cannot yield a definite conclusion since an accidental vanishing 
at that scale cannot be ruled out. 
Yet we have two measures for 
mixing, namely $\Delta M_D$ and $\Delta \Gamma _D$, and their 
oscillatory dependence on $m_c$ in general will be out of phase.

\section{Contributions to 
$D^0 \leftrightarrow \bar D^0$ 
from Exclusive Channels }

We have stated above that the OPE expectation of in particular 
$y_D \sim {\cal O}(10^{-3})$ is highly remarkable since 
{\em a priori} one would estimate it to be an order of 
magnitude larger, see Eq.\,(\ref{GENERAL}). We will illustrate 
this point by considering transitions to two pseudoscalar mesons, which
are common to 
$D^0$ and $\bar D^0$ decays and can thus communicate between 
them:   
\beq 
D^0 \; \; \stackrel{CS\,} \longrightarrow \; \; 
K^+ K^- ,\; 
\pi ^+ \pi ^- \; \; \stackrel{CS\,} \longrightarrow \; \; 
\bar D^0 \; , 
\label{DKK}
\eeq 
\bea  
D^0 &\; \; \stackrel{CA\,} \longrightarrow \; \;& 
K^- \pi ^+ \; \; \stackrel{CS^2} \longrightarrow \; \; 
\bar D^0 
\nonumber 
\\ 
D^0 &\; \; \stackrel{CS^2} \longrightarrow \; \;& 
K^+ \pi ^- \; \; \stackrel{CA\,} \longrightarrow \; \; \bar D^0 
\; . 
\label{DKPI}
\eea 
where $CA$, $CS$ and $CS^2$ denotes the channel as 
Cabibbo allowed, Cabibbo suppressed and doubly Cabibbo 
suppressed, respectively. 

In the $SU(3)$ limit one obviously has 
$\Delta \Gamma (D^0 \to K\bar K,\pi \pi , K\pi , 
\pi \bar K) = 0$ since the amplitudes for 
Eqs.\,(\ref{DKPI}) would then be equal in size and opposite in sign 
to those of Eq.\,(\ref{DKK}).  
Yet the measured branching ratios \cite{PDG98}
\bea 
{\rm BR}(D^0 \to K^+K^-) &=& (4.27 \pm 0.16) \cdot 10^{-3} \\ 
{\rm BR}(D^0 \to \pi ^+ \pi ^-) &=& (1.53 \pm 0.09) \cdot 10^{-3} \\ 
{\rm BR}(D^0 \to K^- \pi ^+) &=& (3.85 \pm 0.09) \cdot 10^{-2} \\ 
{\rm BR}(D^0 \to K^+ \pi ^-) &=& (2.8 \pm 0.9) \cdot 10^{-4}
\eea
show very considerable $SU(3)$ breakings:  
\bea 
\frac{{\rm BR}(D^0 \to K^+K^-)}{{\rm BR}(D^0 \to \pi^+\pi^-)} 
&\simeq& 2.8 \pm 0.2  
\label{KKPIPI}   
\\ 
\frac{{\rm BR}(D^0 \to K^+ \pi ^-)}{{\rm BR}(D^0 \to K^- \pi ^+)} 
&\simeq& (3 \pm 1) \cdot \tan^4 {\theta _C} 
\label{RATIO2B} 
\eea  
compared to ratios of unity and 
$\tan^4{\theta_C}$, 
respectively, in the symmetry limit. 
 
One would then conclude that the 
$K\bar K,\pi \pi , K\pi , 
\pi \bar K$ contributions to $\Delta \Gamma$  
should be merely Cabibbo suppressed with flavor $SU(3)$ 
providing only moderate further reduction -- similar to the 
general expectation of Eq.\,(\ref{GENERAL}): 
\beq 
\left. \frac{\Delta \Gamma }{\Gamma }\right| 
_{D \to K\bar K,\pi \pi , K\pi , 
\pi \bar K} \sim {\cal O}(0.01) \; .
\eeq 
Yet despite these large $SU(3)$ breakings an almost complete 
cancellation
takes place  between their contributions to 
$D^0 \!- \!\bar D^0$ oscillations: 
$$  
{\rm BR}(D^0 \to K^+K^-) + {\rm BR}(D^0 \to \pi ^+ \pi ^-) 
- 2 \sqrt{{\rm BR}(D^0 \to K^- \pi ^+) 
{\rm BR}(D^0 \to K^+ \pi ^-)} \simeq 
$$ 
\beq \left(  - 8 ^{+12}_{-10} \right) \cdot 10^{-4} 
\label{PP}
\eeq
to be compared to 
$$  
{\rm BR}(D^0 \to K^- \pi ^+) + 
{\rm BR}(D^0 \to K^+K^-) + {\rm BR}(D^0 \to \pi ^+ \pi ^-) 
+ {\rm BR}(D^0 \to K^+ \pi ^-) \simeq 
$$ 
\beq 
(4.46 \pm 0.01 ) \cdot 10^{-2} 
\eeq
In principle, a note of caution should be sounded here: In 
writing down Eq.\,(\ref{PP})  
we have ignored the 
possibility that $SU(3)$ {\em breaking} final state 
interactions can generate a strong phase shift 
$\delta_{K\pi}$  
between $D^0 \to K^- \pi ^+$ and $D^0 \to K^+ \pi ^-$ 
amplitudes. If this happens,  
the last interference term 
then gets multiplied by a factor $\cos{\delta _{K\pi}}$. 
Here and in what follows we neglect this phase shift as motivated
by the naive quark level diagrams.\footnote{This fully conforms 
to the spirit 
of the OPE
description whose prediction we are trying to examine at the level of 
hadrons.}
Whether it is small as suggested by some is not clear \cite{WOLF}; 
in any case it could have a significant 
impact on the cancellations among the different terms.  
Yet we meant this discussion only as a qualitative 
illustration of our argument on the relation between $SU(3)$ symmetry 
and duality. 

There is evidence that Eq. (\ref{KKPIPI}) 
overstates the amount of $SU(3)$ breaking in {\em inclusive} 
transitions: the data on Cabibbo suppressed four body 
modes read \cite{PDG98} 
\bea 
\nonumber
{\rm BR}(D^0 \to K^+K^-\pi ^+ \pi ^-) &=& (2.52 \pm 0.24) 
\cdot 10^{-3} \\ 
{\rm BR}(D^0 \to \pi ^+ \pi ^-\pi ^+ \pi ^-) &=& 
(7.4 \pm 0.6) \cdot 10^{-3} \; ; 
\label{BR4B}
\eea 
i.e., again these exclusive channels exhibit very 
sizeable $SU(3)$ breaking 
\beq 
\frac{{\rm BR}(D^0 \to K^+K^-\pi ^+ \pi ^-)}
{{\rm BR}(D^0 \to \pi^+\pi^-\pi ^+ \pi ^-)} 
\simeq 0.34 \pm 0.04\;; 
\label{RATIO4B} 
\eeq
Yet  
adding these two- and four-body modes then 
leads to a result which is quite compatible 
with equality of the combined rates: 
\beq 
\frac{{\rm BR}(D^0 \to K^+K^-, \, K^+K^-\pi ^+ \pi ^-)}
{{\rm BR}(D^0 \to \pi ^+\pi ^-, \; \pi ^+\pi ^-\pi ^+ \pi ^-)} 
\simeq 0.8 \pm 0.1\;.
\label{CAB} 
\eeq  
While this sum cannot be unambiguously related to the violation of
$SU(3)$ in the fully inclusive rates  
$\Gamma (c \to s \bar s u)$ vs. $\Gamma (c \to d \bar d u)$,  
the observed trend is at least suggestive.
\vspace*{.2cm}

To summarize the discussion in this Section:
\begin{itemize}
\item 
The $SU(3)$ breaking in {\em exclusive} nonleptonic 
channels is naturally   
expected to be sizeable, and this is indeed what 
is observed, see Eq. (\ref{KKPIPI}). 
The deviations from the symmetric
case are  actually substantially larger than what had been 
anticipated by most authors. 
\item 
Quark based calculations lead to the prediction that 
{\em inclusive} $D$ decays exhibit $SU(3)$ invariance 
to a high degree, since the symmetry breaking in described 
by $m_s^2/m_c^2 \sim {\cal O}(0.01)$. 
\item 
Emerging data provide the first indication that 
$SU(3)$ breaking is quite reduced when one sums up  
over various nonleptonic channels, see Eq.\,(\ref{CAB}).  
\item 
Likewise the overall contributions to 
$\Delta \Gamma$ from channels with 
two pseudoscalar mesons in the final state appear to be considerably 
reduced, see Eq.\,(\ref{PP}). 
\end{itemize}

\section{Experimental Bounds and Lessons on Duality}

The present experimental landscape
can be  portrayed by the following numbers inferred 
from various analyses of $D^0 \to K^+K^-$ vs. 
$D^0 \to K^- \pi ^+$ and $D^0 \to K^+\pi ^-$ 
vs. $D^0 \to K^- \pi ^+$. From general bounds 
on mixing one can infer: 
\beq 
|x_D| \; , \; |y_D| \; \leq   0.028    \, , \;  
\; 95\, \% \; {\rm C.L.} \qquad {\rm CLEO}\; \cite{ASNER} 
\eeq 
Targeting more specifically width differences 
one finds 
\beq 
 - 0.04 \leq y_D \leq 0.06  \, , \; \; 90\, \% \; {\rm C.L.} 
\qquad {\rm E 791} \;  \cite{E791}
\eeq 
\beq 
- 0.058 \leq y_D^{\prime} \leq 0.01 \, , \;  
\; 95\, \% \; {\rm C.L.} \qquad {\rm CLEO} \; \cite{ASNER}  
\label{CLEOBOUND}
\eeq 
The CLEO study analyzes the time evolution of 
$D^0(t) \to K^+ \pi ^-$ and is thus sensitive
to 
\beq 
y_D^{\prime} = y_D \cos \delta _{K\pi} 
- x_D \sin \delta _{K\pi} 
\eeq
where $\delta _{K\pi}$ denotes the strong phase 
between $D^0 \to K^+ \pi ^-$ and 
$\bar D^0 \to K^+ \pi ^-$. A very recent and still 
preliminary FOCUS study compares the lifetimes for 
$D \to K^+ K^-$ and $D \to K\pi$:    
\beq 
y_D = 0.0342 \pm 0.0139 \pm 0.0074  \qquad \;  
{\rm FOCUS} \; \cite{WISS} 
\label{FOCUS} 
\eeq

At this point we want to summarize and draw the following 
conclusions: 
\begin{itemize}
\item 
Based on general grounds one expects
\beq 
\Delta M_D\; , \; \Delta \Gamma _D 
\sim \; SU(3) \; {\rm breaking} \; 
\times 2 \sin^2{\theta_C} \times \Gamma _D 
\eeq
The observation of large deviations from $SU(3)$ invariance 
in nonleptonic $D$ decays suggests a conservative 
estimate 
\beq 
\frac{\Delta \Gamma _D}{\Gamma _D}\; \leq \; {\rm few} \times 0.01 
\label{141}
\eeq
with $\Delta M_D/\Gamma _D$ being somewhat smaller. 
\item 
Specific {\em dynamical} features have to intervene 
to suppress $D^0 - \bar D^0$ below these levels. Such 
features arise naturally in a quark level treatment 
of $SU(3)$ symmetry breaking as it arises in  
an OPE. Assuming local duality 
one obtains from the OPE the prediction 
\beq 
x_D \; , \; y_D \; \sim \; {\cal O}(10^{-3}) 
\label{142}
\eeq 
without invoking {\em additional\,} long distance contributions. 
The main uncertainty  in this prediction rests in the 
size of the relevant hadronic matrix elements. 
The OPE allows for rather mild $SU(3)$ GIM cancellations. Consequently, for
sufficiently small values of $m_c$ there could be unsuppressed
contributions to the oscillation rate. Yet for the actual charm mass
such contributions should be reasonably suppressed compared to
Eq.\,(\ref{141}) since they emerge in higher orders in
$1/m_c$. Therefore, we would consider the degree of suppression of
$x_D$, $y_D$ below the $1\%$ level as a measure of applicability of
local duality. From this perspective, a stronger suppression of $x_D$
compared to $y_D$ seems the natural situation.

\item 
The data have reached the general bound 
of Eq.\,(\ref{GENERAL}).  
Any further reduction in the experimental bound on $y_D$ 
means that 
$D^0 \!-\! \bar D^0$ oscillations proceed more slowly than can be 
understood on the basis of general selection rules (a ``symmetry level'').
\item 
There is some tentative evidence that inclusive decays might 
exhibit the effective $SU(3)$ invariance expected to arise on 
the quark level. 
\item 
If the suggestion coming from the FOCUS data is confirmed that
actually $y_D \sim {\cal O}(0.01)$ holds then one of two conclusions
can be drawn: Either $\Delta M_D$ is just ``around the corner",
i.e. a moderate improvement in experimental sensitivity should reveal
a nonvanishing value for it without establishing the intervention of
New Physics. This would mean we had seriously underestimated the size
of the relevant matrix elements.  Or it would represent a clear-cut
violation of local quark-hadron duality at the charm scale.

\end{itemize}  
\vspace*{.2cm}
{\it Note added:} After submitting this paper  we were
informed about the publication \cite{georgi} (N.U. is grateful to A.\,Petrov
for bringing our attention to it) where it was first proposed that
contributions due to chiral symmetry breaking that are subleading in
$1/m_c$ can generate a moderately larger value for $\Delta M_D$
than the SM box estimate. That paper focussed on an analogue
of the OPE for the $\Delta C\!=\!1$ transitions rather than directly for
$\Delta C\!=\!2$. Our analysis differs in a number of conceptual as well as
technical aspects.
We analyze the OPE for $\Delta C\!=\!2$ transitions and
obtain the formally leading term linear in $m_s$ that had been missed in
\cite{georgi}. We also address the difference between
$\Delta M_D$ and $\Delta \Gamma _D$ explicitly.

\vspace*{.2cm} 

{\bf Acknowledgments:}~~This work has been supported by the National
Science Foundation under grant number PHY96-05080 and by RFFI grant
\#\,99-02-18355.  We thank D.~Asner and A.~Petrov for helpful
comments, and M.~Lublinsky for discussions. N.U.\ gratefully
acknowledges the hospitality of Physics Department of the Technion
and the support of the Lady Davis grant during completion of this
paper.


\end{document}